\documentclass[journal=jacsat,manuscript=article]{achemso}

\usepackage[version=3]{mhchem} % Formula subscripts using \ce{}
\usepackage[T1]{fontenc}       % Use modern font encodings

\usepackage{graphicx}
\graphicspath{ {./_Figures/} }
\usepackage{epstopdf}
\usepackage{color}         			% color text
\usepackage[normalem]{ulem}			% Allow strikeout formatting of text
\usepackage{soul}

\author{S. L. Zhang}
\affiliation{Clarendon Laboratory, Department of Physics, University of Oxford, Parks Road, Oxford, OX1~3PU, United Kingdom}
\author{A. Bauer}
\affiliation{Physik Department, Technische Universit\"at M\"unchen, 85748 Garching, Germany}
\author{D. M. Burn}
\affiliation{Diamond Light Source, Didcot, OX11~0DE, United Kingdom}
\author{P. Milde}
\affiliation{Institut f\"ur Angewandte Photophysik, TU Dresden, 01069 Dresden, Germany}
\author{E. Neuber}
\affiliation{Institut f\"ur Angewandte Photophysik, TU Dresden, 01069 Dresden, Germany}
\author{L. M. Eng}
\affiliation{Institut f\"ur Angewandte Photophysik, TU Dresden, 01069 Dresden, Germany}
\author{H. Berger}
\affiliation{Crystal Growth Facility, Ecole Polytechnique F\'{e}d\'{e}rale de Lausanne (EPFL), CH-1015 Lausanne, Switzerland}
\author{C. Pfleiderer}
\affiliation{Physik Department, Technische Universit\"at M\"unchen, 85748 Garching, Germany}
\author{G. van~der~Laan}
\affiliation{Magnetic Spectroscopy Group, Diamond Light Source, Didcot, OX11~0DE, United Kingdom}
\author{T. Hesjedal}
\affiliation{Clarendon Laboratory, Department of Physics, University of Oxford, Parks Road, Oxford, OX1~3PU, United Kingdom}
\email{Thorsten.Hesjedal@physics.ox.ac.uk}

\title{Multidomain Skyrmion Lattice State\\ in Cu$_2$OSeO$_3$}


\begin{document}

\newpage
\begin{abstract}
Magnetic skyrmions in chiral magnets are nanoscale, topologically-protected magnetization swirls that are promising candidates for spintronics memory carriers. Therefore, observing and manipulating the skyrmion state on the surface level of the materials are of great importance for future applications. Here, we report a controlled way of creating a multidomain skyrmion state near the surface of a Cu$_{2}$OSeO$_{3}$ single crystal, observed by soft resonant elastic x-ray scattering. This technique is an ideal tool to probe the magnetic order at the $L_{3}$ edge of $3d$ metal compounds giving a depth sensitivity of ${\sim}50$\,nm. The single-domain sixfold-symmetric skyrmion lattice can be broken up into domains overcoming the propagation directions imposed by the cubic anisotropy by applying the magnetic field in directions deviating from the major cubic axes. Our findings open the door to a new way to manipulate and engineer the skyrmion state locally on the surface, or on the level of individual skyrmions, which will enable applications in the future.
\end{abstract}

%%%%%%%%%%%%%%%%%%%%%%%%%%%%%%%%%%%%%%%%%%%%%%%%%%%%%%%%%%%%%%%%%%%%%

Magnetic skyrmions are topologically stable, vortex-like magnetization patterns, and they form periodic lattices in chiral magnets~\cite{Pf_MnSi_Science_09,2010:Jonietz:Science,2010:Yu:Nature,Tokura_CuOSeO_LTEM_Science_12,Rocsh_MnSi_emergent_NatPhys,2013:Milde:Science,2013:Fert:NatureNano,Tokura_review_skyrmion_Natnano_13,Romming2013,Tokura_CuOSeO-MnSi_ratchet_Natmater_14,2015:Schwarze:NatureMater}.
Skyrmions are found in non-centrosymmetric crystals in which the Dzyaloshinskii-Moriya interaction
arises, such as the cubic chiral magnets MnSi, FeGe, or FeCoSi~\cite{Pf_MnSi_Science_09, 2010:Munzer:PhysRevB, Adams_Domains_2010, 2010:Yu:Nature, 2011:Yu:NatureMater, 2012:Tonomura-MnSi, 2013:Yu-MnSi}, which are metallic or semiconducting.
They are attractive objects for information processing applications~\cite{2010:Jonietz:Science, Tokura_review_skyrmion_Natnano_13, 2013:Fert:NatureNano} owing to their topological robustness~\cite{Rocsh_MnSi_emergent_NatPhys, 2013:Milde:Science} and ease of manipulation by spin or magnon currents~\cite{2010:Jonietz:Science,Rocsh_MnSi_emergent_NatPhys,2013:Milde:Science,Tokura_CuOSeO-MnSi_ratchet_Natmater_14,Romming2013}.
An important ingredient for memory applications is the formation of magnetic domains and their manipulation, as demonstrated in the racetrack memory~\cite{2008:Parkin}. This concept has also been proposed for magnetic skyrmion systems~\cite{2013:Fert:NatureNano,Tokura_review_skyrmion_Natnano_13,2015:Zhang-Topo,2015:Liang-MnSi-wires}. However, the skyrmion state in cubic chiral magnets are usually a single-domain, uniform vortex lattice, which does not support the manipulation of the individual skyrmion. Therefore, the break-up of the single-domain skyrmion lattice into a multidomain state, in analogy to the magnetic domains in ferromagnetic media, is pivotal for skyrmion-based memory applications.
So far, only double-split, non-dynamic skyrmion domains have been observed in FeCoSi~\cite{2010:Munzer:PhysRevB, Adams_Domains_2010, White_Cu2OSeO3_LTEM_PNAS_15}.
However, little is known about the occurrence of skyrmion domains in other systems, and controlled ways of their manipulation.
On the other hand, insulating magnetoelectric materials, such as Cu$_2$OSeO$_3$~\cite{Thomas_Cu2OSeO3_neutron_PRB_08, Tokura_CuOSeO_LTEM_Science_12, Tokura_CuOSeO_FE_PRB_12, Fudan_Cu2OSeO3_DFT_PRL_12, Berger_Cu2OSeO3_phase_diagram_Srep_15},
provide additional degrees of freedom for the manipulation of skyrmions~\cite{White_Cu2OSeO3_E_rotation_IOP_12, White_Cu2OSeO3_E_rotation_PRL_14, 2014:Lin:PhysRevLett}.

Cu$_{2}$OSeO$_{3}$ is a local-moment insulator and belongs to the group of cubic chiral magnets with space group $P2_13$~\cite{1985:Effenberger:MonatshChem}. Its magnetism is governed by a hierarchy of energy scales that can be described by a Ginzburg-Landau approach also accounting for thermal fluctuations~\cite{Pf_MnSi_Science_09, Dmitrienko_Cu2OSeO3_theory_JMMM_15}. On the strongest scale, the combination of ferromagnetic and antiferromagnetic exchange interactions effectively leads to a ferrimagnetic 3-up-1-down spin configuration of the Cu$^{2+}$ moments on two types of nonequivalent sites in the unit cell~\cite{1977:Kohn:JPhysSocJpn, Berger_Cu2OSeO3_NMR_PRB_10}. On the intermediate scale, the Dzyaloshinskii-Moriya interaction arises as the leading-order spin-orbit coupling term, resulting in a long-wavelength incommensurate helical modulation with the periodicity of $\lambda_{\mathrm{h}} \approx 62$\,nm~\cite{Tokura_CuOSeO_LTEM_Science_12, Pf_CuOSeO_PRL_12, Tokura_CuOSeO_rotation_PRB_12}. On the weakest energy scale, higher-order spin-orbit coupling terms give rise to a cubic anisotropy, favoring helical propagation along the crystalline $\langle100\rangle$ axes. After zero-field cooling below the transition temperature of $T_\mathrm{C} = 58$\,K, multiple macroscopic helical domains oriented along the three equivalent $\langle100\rangle$ directions form. In magnetic fields larger than $H_{c1}$, the propagation vector of the helix aligns along the field direction, giving rise to a single-domain conical state. With increasing field, the spins increasingly tilt towards the field direction until entering a field-polarized regime above the second critical field $\mu_{0}H_{c2} \approx 0.1$\,T. In a phase pocket in small fields just below $T_\mathrm{C}$, a hexagonal skyrmion lattice is observed in the plane perpendicular to the applied field, forming tubes along the field direction~\cite{Pf_MnSi_Science_09, Tokura_CuOSeO_LTEM_Science_12}. 

Commonly, the skyrmion lattice is a single-domain, long-range ordered state~\cite{2011:Adams:PhysRevLett,Pf_CuOSeO_PRL_12}. In reciprocal space, the skyrmion lattice may be described by three basis vectors $\boldsymbol{\tau}_{1}$, $\boldsymbol{\tau}_{2}$, and $\boldsymbol{\tau}_{3}$ that lie in the plane perpendicular to the magnetic field direction and are separated by $60^{\circ}$~\cite{Rocsh_MnSi_emergent_NatPhys}. The absolute value $\tau$ corresponds to the `lattice constant' of the skyrmion lattice, i.e., the core-to-core distance $a$, with $a = 2 \pi / (\sqrt{3}\tau)$. Due to the cubic anisotropy, one of the three $\tau$ vectors is oriented along a $\langle$100$\rangle$ or $\langle$110$\rangle$ direction -- if the latter is available in the skyrmion lattice plane~\cite{Pf_CuOSeO_PRL_12, White_Cu2OSeO3_E_rotation_PRL_14}. A reorientation from $\langle$110$\rangle$ to $\langle$100$\rangle$ as a function of increasing temperature and field was reported in Cu$_{2}$OSeO$_{3}$, suggesting a delicate balance of the weak cubic anisotropies~\cite{Tokura_CuOSeO_rotation_PRB_12}.

In this Letter, we performed systematic resonant elastic x-ray scattering (REXS) studies in reflection geometry on a Cu$_{2}$OSeO$_{3}$ single crystal. 
We unambiguously identified the helical, conical, and skyrmion lattice states. 
For magnetic fields tilted away from the crystalline [001] axis, the scattering pattern of the skyrmion lattice breaks up into a multitude of sixfold-symmetric sets which are, in general, not aligned along major crystallographic axes. Thoroughly performed photon energy scans reveal identical energy dependencies of all magnetic peaks. 
In a recent study, Langner \textit{et al.}\ found two slightly rotated, sixfold-symmetric scattering patterns that differ in photon energy dependency, which are claimed to be associated with the two different Cu sites~\cite{Tokura_CuOSeO_REXS_PRL_14}.
As will be shown, the multidomain skyrmion state clearly has a fundamentally different origin. It is explained in the framework of competing anisotropies and magnetoelectric effects.

\begin{figure}
\includegraphics[width=8.46cm]{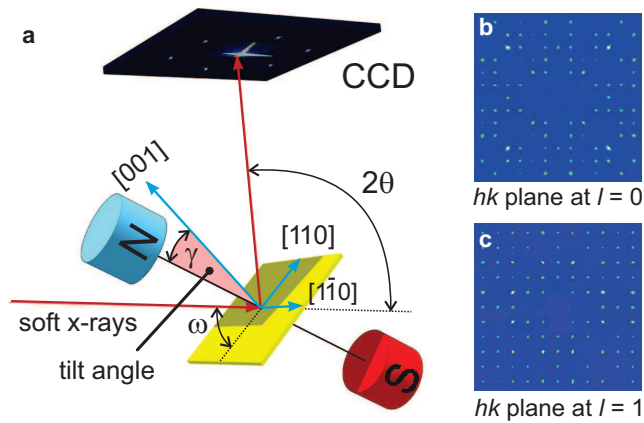}
\caption{REXS setup and single-crystal x-ray diffraction. (a) Scattering geometry and field configuration in RASOR. The magnetic field is provided by permanent magnets in a rotatable variable field assembly. (b,c)~Single-crystal off-resonance diffraction pattern showing the reciprocal space points in the $hk$ plane for $l$=0 and $l$=1, for which the (0,0,1) reflection is forbidden.}
\label{fig_1}
\end{figure}

REXS is a powerful, surface-sensitive technique that allows for simultaneously resolving long-wavelength modulations along all three directions in reciprocal space. It combines element-selectivity with extremely fast time resolution. Our experiment was carried out on beamline I10 at the Diamond Light Source, UK, in the ultrahigh vacuum soft x-ray diffractometer RASOR~\cite{RASOR_review}. Figure~\ref{fig_1}a illustrates the geometry of the setup. The horizontally polarized soft x-rays are tuned to the Cu $L_{3}$ edge at a photon energy of 931.25\,eV. The scattered light is captured by either a CCD camera or a photodiode at an angle $2\theta$ with respect to the incoming beam. Typically, we used a beam size of $\sim$100$\times$100~$\mu$m$^2$ and an exposure time of 2\,ms for each single image. A carefully polished Cu$_{2}$OSeO$_{3}$ high-quality single-crystal measuring $5\times3\times0.5\,\mathrm{mm}^{3}$, with the edges along $[110]$, $[1\bar{1}0]$, and [001], was positioned as sketched. 
The sample was confirmed to be of high quality by means of single-crystal x-ray diffraction on a Rigaku SuperNova using a Mo $K_{\alpha 1}$ source, see Figure \ref{fig_1}b,c. It is of uniform lattice chirality as determined by electron backscatter diffraction (EBSD).
For measurements on RASOR, the magnetic field direction is typically fixed with respect to the sample plane, i.e., both the magnets and the sample are moved together by the goniometer angle $\omega$. However, the magnets can also be rotated away from the surface normal direction in the scattering plane by a tilt angle $\gamma$, where $\gamma$=$0^{\circ}$ corresponds to the field along the $[001]$ axis of the sample.

When probing a skyrmion lattice, x-rays acquire the momentum transfer directly from the periodicities of the vortex ordering. Therefore, the scattering form factors for the helical and skyrmion orders are significantly different in REXS --- unlike for small angle neutron scattering (SANS) where they are almost the same.
For REXS, the magnetic modulation wave vectors appear as the satellite peaks coupled to the (0,0,1) structural peak.
For our measurements on Cu$_{2}$OSeO$_{3}$, we choose the resonant condition at the Cu $L_{3}$ edge and the (0,0,1) peak is the only accessible reciprocal lattice point for the corresponding x-ray wavelength of 1.33\,nm at large scattering angles (${\sim}96.5^{\circ}$). Using a distance between sample and detector of only ${\sim}120$\,mm, our REXS setup is sensitive to all three components of the magnetic reciprocal space vectors, whereas SANS only detects the two components perpendicular to the incoming neutron beam.
Also, the x-ray attenuation length at the Cu $L_{3}$ peak maximum is 94\,nm, which for the chosen geometry gives a sampling depth of 34\,nm corresponding to 38 unit cells. 
Hence, REXS only probes a depth that is comparable to the helical wavelength, and is ideal for studying near-surface magnetic structures.
It has to be mentioned that the influence of the surface on the complex magnetic texture of the skyrmion lattice remains an open question. The surface, as a boundary of abruptly changing properties, may induce extraordinary electronic~\cite{Review_TI} or magnetic~\cite{Hamburg_Ir/Fe} properties in strongly correlated electron systems, and similar effects may also be expected for $P2_{1}3$ systems~\cite{Monchesky_Karhu_sim_PRB_12, Monchesky_Wilson_unwind_PRB_13, Monchesky_MnSi_surface_twist_PRB_14}. Commonly employed techniques, such as SANS or Lorentz transmission electron microscopy, are insensitive to surface effects.
Finally, our REXS experiment allows for very short exposure times on the order of one millisecond and video rate dynamic imaging.

\begin{figure}
\includegraphics[width=8.46cm]{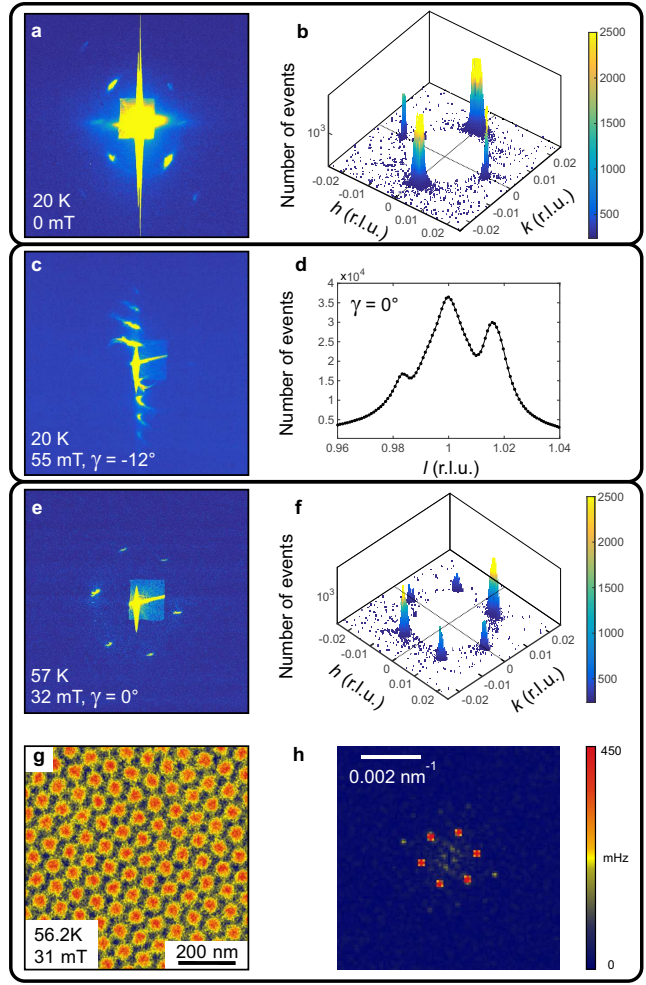}
\caption{Typical REXS data of the different magnetic phases.
(a)~Summed CCD images in the helical state. (b)~Reciprocal space map of the $hk$ plane at $l$=1, obtained from processing the CCD image (a). (c)~Summed images in the conical state at $\gamma$=$-12^{\circ}$. (d)~Cut through reciprocal space along $l$ for $\gamma$=$0^{\circ}$. (e)~Sum of images in the skyrmion lattice state. (f)~Reciprocal space map of the $hk$ plane at $l$=1, obtained from processing the CCD image (e).
(g)~MFM image of the skyrmion lattice obtained in an out-of-plane field of 31~mT. The Fourier transform (h) reveals the hexagonal pattern in agreement with the REXS data.
}
\label{fig_2}
\end{figure}

Figure~\ref{fig_2} shows representative REXS data for the different magnetic phases of Cu$_2$OSeO$_3$. In order to acquire the magnetic satellites, we collected reciprocal space maps~(RSMs) around the (0,0,1) peak. At fixed $\gamma$, RSMs are obtained by rocking $\omega$ by $\pm 2.6^{\circ}$ in $0.05^{\circ}$ steps around the (0,0,1) diffraction condition, while keeping the CCD detector angle coupled. This results in a series of CCD images of these $\omega$-2$\theta$ scans. Adding together such a series yields an integrated diffraction pattern that reveals the symmetry of the magnetic satellites relative to the (0,0,1) peak. However, it does not directly provide quantitative information in reciprocal space. The detailed experimental methods and explanations can be found in the Supporting Information. The large intensity maximum at the center of each image is the (0,0,1) diffraction peak, while the vertical streak is ascribed to the blooming of the CCD pixels. The observed (0,0,1) peak is rather strong, despite being crystallographically forbidden for space group $P2_13$ (see Figure \ref{fig_1}b,c). We attribute this effect to Templeton scattering, where the anisotropic third-rank tensor stemming from the mixed dipole-quadrupole term allows for the extinction peak to appear for non-centrosymmetric crystals at the x-ray resonance condition~\cite{Templeton_PRB_94}.

Figure~\ref{fig_2}a shows the sum of CCD images in zero field (without the magnets mounted). Around the central (0,0,1) peak we observe four clear magnetic satellites that correspond to the helical state, together with weak maxima arising from higher-order or double scattering. The map of the $hk$ plane at $l=1$ is shown in Figure \ref{fig_2}b where the (0,0,1) peak is manually removed in order to increase the contrast. Satellites at ({$\pm$} {$q_{\mathrm{h}}$},0,1), (0,{$\pm$}{$q_{\mathrm{h}}$},1), and (0,0,1{$\pm$} {$q_{\mathrm{h}}$}) (not shown) with {$q_{\mathrm{h}}$}=(0.016$\pm$0.002)\,r.l.u.\, correspond to 55-56.5\,nm in real space and easy $\langle100\rangle$ axes for the helical propagation, in agreement with SANS data~\cite{Pf_CuOSeO_PRL_12}. The slightly reduced helix wavelength may be associated with surface effects.

In the conical state, the propagation vector is aligned along the field direction, and for $\gamma$=$0^{\circ}$ the resulting satellites at (0,0,1{$\pm$}{$q_{\mathrm{h}}$}) are not visible in the integrated image from the coupled $\omega$-2$\theta$ scans. Tilting the field, however, leads to a tilt of the conical propagation vectors relative to the $l$-direction. The resulting sum over CCD images is shown in Figure \ref{fig_2}c for $\gamma$=$-12^{\circ}$, where a chain of vertically aligned maxima is characteristic of pronounced higher-order scattering. This has not been observed before and may be attributed to anharmonicities that are induced at the surface of the sample.
A cut through reciprocal space along $l$ for $\gamma$=$0{^\circ}$, see Figure \ref{fig_2}d, reveals that the (0,0,1) peak is indeed surrounded by the conical satellites at $\pm q_{\mathrm{h}}$. The asymmetry of the intensity is due to the diffuse scattering from the lattice reflection.
 
In a field of 32\,mT at 57\,K, we observe the characteristic sixfold diffraction pattern of the skyrmion lattice, see Figure \ref{fig_2}e. The RSM shown in Figure \ref{fig_2}f reveals that the three basis vectors have the same size as the helical wavevector, are rotated by $60^{\circ}$, where one of them aligns along the $h$ direction. The considerable difference between the scattering intensities of the helical, conical, and lattice skyrmion phases suggests that the form factors, structure factors, as well as the scattering amplitudes are characteristic for each of these spin textures. This finding highlights a main advantage of REXS in that it directly probes magnetic order, and thus skyrmion order, as compared to SANS.

Complementary magnetic force microscopy (MFM) was performed to obtain real-space images of the skyrmion lattice state. Figure \ref{fig_2}g shows an MFM image in the skyrmion phase in an applied out-of-plane field of 31\,mT at a temperature of 56.2~K. The MFM image has been background-corrected and high-pass filtered. The color scale covers a range of 450\,mHz. The fast Fourier transform, shown in Figure \ref{fig_2}h, is computed before the filtering step and thus includes all spatial frequencies. The sixfold pattern is in excellent agreement with the diffraction result shown in Figure \ref{fig_2}e,f.

\begin{figure}
\includegraphics[width=8.46cm]{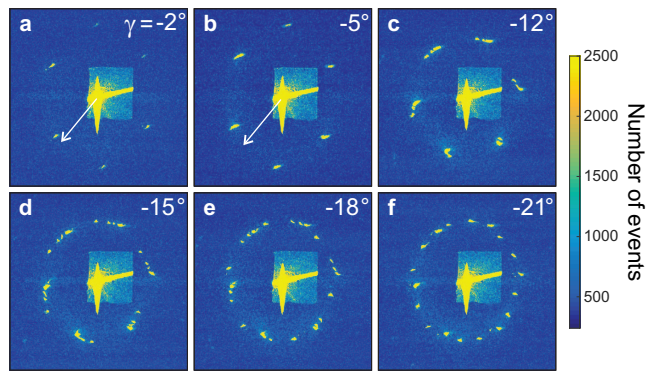}
\caption{Sum of CCD images in the skyrmion lattice phase at different field angles $\gamma$ (a-f). The sample was field-cooled in 32\,mT at the indicated $\gamma$ from 65\,K down to 56.5\,K before a RSM was recorded. As a function of increasing $\gamma$ the single sixfold scattering pattern prevalently splits into several sixfold symmetric subsets. The approximate ($\tau$,0,1) direction is indicated by the white arrow.}
\label{fig_3}
\end{figure}

Next, we focus on the skyrmion lattice where Figure \ref{fig_3} presents sums of CCD images for different magnetic field orientations $\gamma$. Starting well above $T_\mathrm{C}$ at 65\,K, a field of 32\,mT is applied under a fixed $\gamma$. After cooling the sample to 56.5\,K and stabilizing the temperature for 15\,min, a RSM is carried out. After the scan, the sample is heated to 65\,K. Up to a tilt of $\gamma$=$-2^{\circ}$, see Figure \ref{fig_3}a, the sixfold scattering pattern remains unchanged with one of the basis vectors essentially aligned along $h$. For a further increased $\gamma$, see Figure \ref{fig_3}b, the sixfold pattern reorients. Note that for $\gamma$=$-5^{\circ}$ the skyrmion plane is tilted by ${\sim}5^{\circ}$ with respect to the crystalline [001]. As a result, no easy axes, as defined by the cubic anisotropy, are available for the propagation vectors perpendicular to the field direction. Therefore, the propagation vectors only roughly follow the projected easy axes directions.

As shown in Figure \ref{fig_3}c--f, for even larger values of $\gamma$, the sixfold intensity distribution splits up into several sixfold subsets that become increasingly pronounced with increasing $\gamma$. For $\gamma$=$-21^{\circ}$, the largest value accessible, the diffraction pattern finally resembles a beaded necklace. The intensity distributions can be reproduced by repeating the same temperature-field history. Positive as well as negative values of $\gamma$ lead to the same results. As a function of time, the intensity distributions stay unchanged for at least one hour. Moreover, once a split pattern has formed, the rotation of $\gamma$ back to $0^{\circ}$ at constant temperature and magnetic field does not alter the pattern (on a timescale of at least 15\,min).

\begin{figure}
\includegraphics[width=8.46cm]{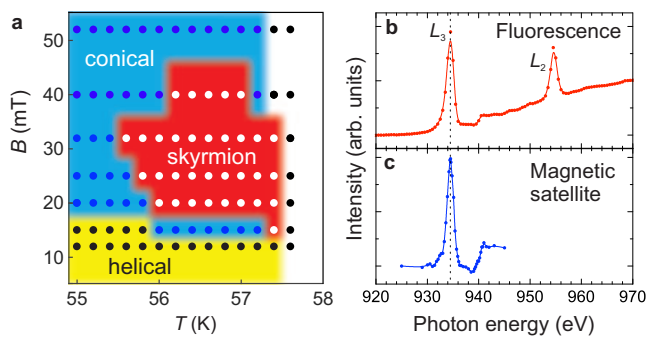}
\caption{(a) Magnetic phase diagram mapped by REXS for the field along [001] ($\gamma$=$0^{\circ}$). (b) X-rays absorption spectrum obtained in fluorescence-yield mode and (c)~magnetic intensity spectrum as a function of photon energy for the skyrmion lattice state at 57\,K and 32\,mT ($\gamma$=$0^{\circ}$). The experimental procedure is described in the Supporting Information.}
\label{fig_4}
\end{figure}

The magnetic phase diagram extracted from the REXS measurements is depicted in Figure \ref{fig_4}a. It is consistent with SANS and magnetometry data, taking into account the different cryostats and sample shapes used~\cite{Pf_CuOSeO_PRL_12}. In order to further elaborate on the complex scattering pattern in the skyrmion lattice phase, we performed photon energy scans in fluorescence-yield mode (Figure \ref{fig_4}b) and for the magnetic satellites (Fig.~\ref{fig_4}c). In stark contrast to Ref.\ \cite{Tokura_CuOSeO_REXS_PRL_14}, no splitting of the peak at the $L_{3}$ edge is observed in either of these spectra. The absence of splitting is to be expected as the Cu$^\mathrm{I}$ and Cu$^\mathrm{II}$ sites have essentially the same valence charge.\cite{Thomas_Cu2OSeO3_neutron_PRB_08} This fact by itself distinguishes our findings from the scenario reported in Ref.\ \cite{Tokura_CuOSeO_REXS_PRL_14}, where the formation of two independent, misaligned skyrmion lattices on the two Cu sites leads to a moir\'{e} pattern. Instead, in particular supported by the fact that clearly more than two sixfold subpatterns are observed at large field angles $\gamma$, we attribute the complex intensity distribution to a multidomain skyrmion lattice state.

Note that multidomain skyrmion patterns have been observed as the ground state in thinned-down Cu$_{2}$OSeO$_{3}$ bulk samples imaged by Lorentz transmission electron microscopy (LTEM).\cite{Tokura_CuOSeO_rotation_PRB_12, Tokura_CuOSeO-MnSi_ratchet_Natmater_14, White_Cu2OSeO3_LTEM_PNAS_15}
However, while Langner \textit{et al.}\ and we studied the surface of pristine bulk crystals by REXS, LTEM requires thinned-down plates which have undergone preparation steps, altering the experimental conditions slightly, as pointed out by Rajeswari \textit{et al.}\cite{White_Cu2OSeO3_LTEM_PNAS_15} In this case disorder will strongly affect the skyrmion state, and provide an energy landscape due to defects that favors the formation of domains.
Another possible source of the discrepancy between reciprocal and real-space imaging is the fact that the electron beam itself may introduce instabilities in the skyrmion lattice as observed by LTEM.
In contrast to Rajeswari \textit{et al.}'s work, the multidomain pattern reported here does not evolve over time, using an exposure time of only 2\,ms. Therefore, we also exclude a dynamic origin that was introduced as the explanation for a double-split pattern observed in LTEM for long, averaging exposure times of 100\,ms.\cite{White_Cu2OSeO3_LTEM_PNAS_15}

\begin{figure}
\includegraphics[width=8.46cm]{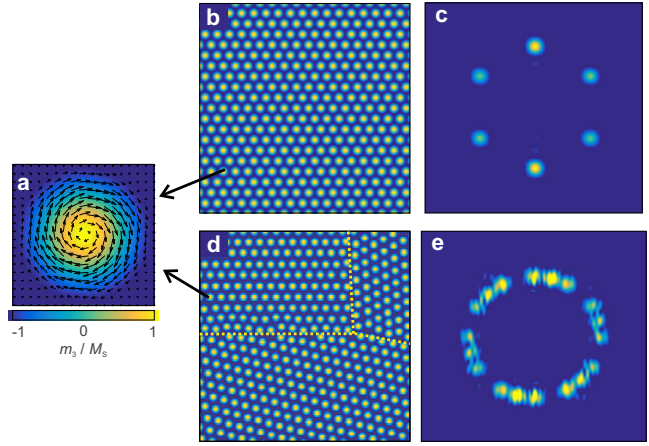}
\caption{Numerical calculations of the REXS signal for different skyrmion lattice states. (a) Magnetization configuration of an individual skyrmion that is used as the motif to construct the lattice for the calculations. The color scale represents the $z$-component of the magnetization unit vector. (b) Single domain skyrmion lattice used for the simulation. (c) Reciprocal space REXS pattern for horizontally polarized x-rays at the Cu $L_3$ edge. (d) Real-space multidomain skyrmion state. (e) Simulated REXS pattern using the same incident x-rays as in (c).
}
\label{fig_5}
\end{figure}

The REXS patterns for different states of the skyrmion lattice, e.g., single versus multidomain, can be expected to be different if the domains are smaller than the area sampled by the x-rays. Figure \ref{fig_5}a shows the magnetization configuration for an individual skyrmion vortex as the motif of the hexagonal lattice in the single-domain skyrmion phase (shown in Figure \ref{fig_5}b). The simulated REXS pattern, assuming horizontally polarized x-rays with the photon energy tuned to the Cu $L_3$ edge, is shown in Figure \ref{fig_5}c. Note that the different scattering intensities for the six magnetic peaks are due to the x-ray polarization dependence~\cite{Blume_1988,2008:vdL}. The simulated pattern is in excellent agreement with our experimental results (see e.g. Figure \ref{fig_2}f). 
Next, we analyze a skyrmion lattice state with three domains. The domain boundaries are indicated by dashed yellow lines (Figure \ref{fig_5}d).
Within each domain a well-defined hexagonal skyrmion lattice is found that is rotated with respect to the neighboring domains.
The simulated diffraction result is shown in Figure \ref{fig_5}e, which recovers the `necklace' pattern as experimentally observed in Figure \ref{fig_3}. 
Note that the formation of the skyrmion domains along the vertical [001] direction can also give rise to the `necklace' diffraction pattern as shown in Figure \ref{fig_3}c-f, in which the three-dimensional skyrmion `tubes' break up into several layers that are differently oriented. However, this scenario will generate great magnetization discontinuities and artificial topological defects that cost much more energy, leading to an extremely unstable state. Therefore, we rule out this possibility, and interpret the multidomain state as only the lateral skyrmion domain formation. The in-plane orientation of the domains shows no clear preferred direction and the fragmentation is far more pronounced than in other chiral magnets.\cite{Adams_Domains_2010} We note that x-ray beam sizes between $50\times50$~$\mu$m$^2$ and $200 \times 200$~$\mu$m$^{2}$ result in similar scattering patterns suggesting an average domain size well below 50~$\mu$m. Significantly, the condensation of these domains can be reproducibly induced using a tilted magnetic fields as the tuning parameter.

We will now discuss the possible origins of the formation of a multidomain skyrmion lattice state. As discussed above, REXS only probes the magnetic ordering on the surface and surface-near areas of the material. Note that the magnetic state on the surface may be quite different from the bulk due to the influence of surface anisotropy. For the analysis, we use the well-established phenomenological model that satisfies the continuum approximation.\cite{Pf_MnSi_Science_09, Bogdanov_IOP_2011} In this model, the total energy of the system is written as a functional $E=\int w({\boldsymbol{m}})dV$, with the energy density $w$ defined in terms of the continuously varying magnetization 
\begin{equation}
w({\boldsymbol{m}})=A(\nabla {\boldsymbol{m}})^2 + D {\boldsymbol{m}}\cdot (\nabla \times  {\boldsymbol{m}}) -  {\boldsymbol{B}}\cdot  {\boldsymbol{m}} + w_A \,\,,
\label{eq_1}
\end{equation}
where $A$ is the exchange stiffness, $D$ is the Dzyaloshinskii-Moriya constant, ${\boldsymbol{B}}$ the external magnetic field, and $w_A$ the anisotropy term.
The additional magneto-electrical coupling term can be neglected in the absence of an external electric field.\cite{White_Cu2OSeO3_E_rotation_PRL_14}
For ${\boldsymbol{B}=0}$, the helical ground state minimizes the total energy, and the magnetization can be written as $ {\boldsymbol{m}}( {\boldsymbol{r}})= {\boldsymbol{m}}({\boldsymbol{q}})e^{i {\boldsymbol{q}}\cdot  {\boldsymbol{r}}} + \mathrm{c.c.}$, where $\boldsymbol{m}(\boldsymbol{q})$ is the Fourier amplitude at $\boldsymbol{q}$, while $\boldsymbol{q}$ is the helix propagation wave vector with a length of $q \approx 0.016$~r.l.u., and $\mathrm{c.c.}$ denotes the complex conjugate. At finite fields, the skyrmion lattice solution takes the form of 
\begin{equation}
\boldsymbol{m}(\boldsymbol{r}) = \frac{1}{3} \sum_{i=1}^{3} 
\left[ \boldsymbol{m} (\boldsymbol{q}_i) e^{i\boldsymbol{q}_i \cdot \boldsymbol{r}} + \mathrm{c.c.} \right] + \boldsymbol{m}_{\text{net}} \,\,,
\label{eq_2}
\end{equation}
where $\boldsymbol{q}_1$, $\boldsymbol{q}_2$, and $\boldsymbol{q}_3$ coherently propagate in the plane perpendicular to $\boldsymbol{B}$, taking the form $\boldsymbol{q}_i=q[\hat{q}_x\text{cos}(\Phi_i)+\hat{q}_y\text{sin}(\Phi_i)]$. $\hat{q}_x$ and $\hat{q}_y$ are the orthogonal unit vectors of the two-dimensional reciprocal space; $\Phi_1$, $\Phi_2$, and $\Phi_3$ describe the azimuthal angles, which are $120^\circ$ apart from each other, and $\boldsymbol{m}_{\text{net}}$ is used to describe the net magnetization in the skyrmion state. 

The azimuthal angles $\Phi_i$ only depend on the sixth-order cubic anisotropy term,\cite{Pf_MnSi_Science_09, White_Cu2OSeO3_E_rotation_PRL_14} and read
\begin{equation}
\begin{split}
w_A^{\text{cubic}(6)} = am_1^2m_2^2m_3^2 + b(m_1^2m_2^4+m_2^2m_3^4+m_3^2m_1^4) \\
+ c(m_2^2m_1^4+m_3^2m_2^4+m_1^2m_3^4) + d(m_1^6+m_2^6+m_3^6) \,\,,
\end{split}
\label{eq_3}
\end{equation}
where $m_1,m_2,m_3$ are the three components of $\boldsymbol{m}$, and $a$, $b$, $c$, and $d$ are amplitudes.
By substituting this into Eq.\ (\ref{eq_1}) the minimum energy is obtained when one of the three $\Phi_i$ is pinned along a $\langle$100$\rangle$ direction, while being a solution of Eq.\ (\ref{eq_2}) for the skyrmion lattice.
We consider a surface anisotropy different from the bulk anisotropy, which extends up to a few unit cells in depth and which may be induced by symmetry breaking.
We assume that across the depth of a few unit cells, the expression simplifies to
\begin{equation}
w_A^{\text{surface}} = K_\text{u}m_3^2 \,\,.
\label{eq_4}
\end{equation}
The uniaxial anisotropy constant $K_\text{u}$ is positive, describing an easy-plane anisotropy, similar to the case of MnSi thin films.\cite{Monchesky_Karhu_sim_PRB_12, Monchesky_MnSi_surface_twist_PRB_14}
By inserting Eq.\ (\ref{eq_4}) into Eq.\ (\ref{eq_1}), one obtains that the energy term possesses $SO(2)$ symmetry, the group of rotations about the $\boldsymbol{B}$ direction. Therefore, the $\Phi_i$ are not pinned, suggesting that all $\boldsymbol{q}_i$ propagation directions in the plane perpendicular to $\boldsymbol{B}$ are degenerate. 

In our REXS experiment, soft x-rays probe $\sim$38 unit cells, i.e., much deeper than the extend of the surface anisotropy. 
At this depth, the system is governed by the competition between $w_A^{\text{cubic}}$ and $w_A^{\text{surface}}$.
For $\boldsymbol{B}\parallel [001]$, the cubic anisotropy dominates over the surface anisotropy and locks $\Phi_i$, as observed in Figure \ref{fig_2}e. When tilting the field $\boldsymbol{B}$ by an angle $\gamma$ an in-plane component of the field arises, inducing a depinning of the skyrmion lattice from the original $\langle$100$\rangle$ direction.
With increasing $\gamma$ the depinning of the skyrmion lattice allows the surface anisotropy to become dominant, therefore the propagation direction becomes arbitrary. Eventually a multidomain skyrmion state is formed, as observed in Figure \ref{fig_3}. 

Furthermore, another important ingredient that may be responsible for the multidomain skyrmion state is the magnetoelectric coupling that separates Cu$_{2}$OSeO$_{3}$ from other cubic chiral magnets. Calculations suggest that a magnetic field along $\langle100\rangle$ results in low overall values of the local electric polarization,\cite{Tokura_CuOSeO_FE_PRB_12} consistent with a single-domain skyrmion lattice for fields along [001]. On the other hand, magnetic fields along other directions lead to large in-plane or out-of-plane electric dipole moments. The complex interplay of the magnetic and electric dipole moments, in combination with the delicate cubic anisotropy \cite{Tokura_CuOSeO_rotation_PRB_12}, may finally induce the formation of multiple domains. This assumption is also corroborated by the reorientation of the skyrmion lattice in external electrical fields.\cite{White_Cu2OSeO3_E_rotation_IOP_12}
In order to distinguish between these contributions, further studies focusing on the magnetoelectric coupling in Cu$_{2}$OSeO$_{3}$ and on a systematic comparison of bulk and surface properties are required. In particular, it is highly desirable to measure the real-space domain structure in rotated magnetic fields by MFM complementing the reciprocal space mapping.

In conclusion, we have fully characterized the magnetic phase diagram of the chiral magnet Cu$_{2}$OSeO$_{3}$ using surface-sensitive REXS, highlighting the potential of this x-ray technique for the study of complex magnetic textures.
More significantly, for the skyrmion lattice, we reproducibly induced the condensation of multiple domains with different in-plane orientations by tilting the magnetic field, offering a new tuning parameter for the manipulation of this topologically non-trivial state.
Note that for device applications it is essential to be able to provide an all-electric way to manipulate the domain state, which is, in principle, achievable in magnetoelectric materials such as Cu$_2$OSeO$_3$. 
Compared to a rigid skyrmion lattice state, the multidomain state will allow for the encoding of information in individual skyrmion domains --- a prerequisite for skyrmion-based memory applications.

%%%%%%%%%%%%%%%%%%%%%%%%%%%%%%%%%%%%%%%%%%%%%%%%%%%%%%%%%%%%%%%%%%%%%
\begin{acknowledgement}

The REXS experiments were carried out on beamline I10 at the Diamond Light Source, UK, under proposals SI-11784 and SI-12958.
S.~L.~Z.\ and T.~H.\ acknowledge financial support by the Semiconductor Research Corporation.
A.~B.\ and C.~P.\ acknowledge financial support through DFG TRR80 and ERC AdG (291079, TOPFIT).
P.~M., E.~N., and L.~E.\ acknowledge financial support through DFG SFB1143.
We thank S.\ Speller (Oxford Materials) for the EBSD measurements and S.\ Komineas for fruitful discussions.

\end{acknowledgement}

\section*{Author Contributions}
S.L.Z., C.P., G.v.d.L. and T.H.\ designed the experiment.
S.L.Z., A.B., D.M.B., G.v.d.L.\ and T.H.\ carried out the resonant elastic x-ray scattering experiments and analyzed the data. 
P.M., E.N., and L.E.\ performed the MFM experiments.
H.B. synthesized the bulk samples.
S.L.Z., A.B., C.P., G.v.d.L.\ and T.H.\ wrote the paper with input and comments from all authors.


\providecommand{\latin}[1]{#1}
\providecommand*\mcitethebibliography{\thebibliography}
\csname @ifundefined\endcsname{endmcitethebibliography}
  {\let\endmcitethebibliography\endthebibliography}{}


\begin{mcitethebibliography}{46}
\providecommand*\natexlab[1]{#1}
\providecommand*\mciteSetBstSublistMode[1]{}
\providecommand*\mciteSetBstMaxWidthForm[2]{}
\providecommand*\mciteBstWouldAddEndPuncttrue
  {\def\EndOfBibitem{\unskip.}}
\providecommand*\mciteBstWouldAddEndPunctfalse
  {\let\EndOfBibitem\relax}
\providecommand*\mciteSetBstMidEndSepPunct[3]{}
\providecommand*\mciteSetBstSublistLabelBeginEnd[3]{}
\providecommand*\EndOfBibitem{}
\mciteSetBstSublistMode{f}
\mciteSetBstMaxWidthForm{subitem}{(\alph{mcitesubitemcount})}
\mciteSetBstSublistLabelBeginEnd
  {\mcitemaxwidthsubitemform\space}
  {\relax}
  {\relax}

\bibitem[M{\"u}hlbauer \latin{et~al.}(2009)M{\"u}hlbauer, Binz, Jonietz,
  Pfleiderer, Rosch, Neubauer, Georgii, and B{\"o}ni]{Pf_MnSi_Science_09}
M{\"u}hlbauer,~S.; Binz,~B.; Jonietz,~F.; Pfleiderer,~C.; Rosch,~A.;
  Neubauer,~A.; Georgii,~R.; B{\"o}ni,~P. \emph{Science} \textbf{2009},
  \emph{323}, 915\relax
\mciteBstWouldAddEndPuncttrue
\mciteSetBstMidEndSepPunct{\mcitedefaultmidpunct}
{\mcitedefaultendpunct}{\mcitedefaultseppunct}\relax
\EndOfBibitem
\bibitem[Jonietz \latin{et~al.}(2010)Jonietz, M\"{u}hlbauer, Pfleiderer,
  Neubauer, M\"{u}nzer, Bauer, Adams, Georgii, B\"{o}ni, Duine, Everschor,
  Garst, and Rosch]{2010:Jonietz:Science}
Jonietz,~F.; M\"{u}hlbauer,~S.; Pfleiderer,~C.; Neubauer,~A.; M\"{u}nzer,~W.;
  Bauer,~A.; Adams,~T.; Georgii,~R.; B\"{o}ni,~P.; Duine,~R.~A.; Everschor,~K.;
  Garst,~M.; Rosch,~A. \emph{Science} \textbf{2010}, \emph{330},
  1648--1651\relax
\mciteBstWouldAddEndPuncttrue
\mciteSetBstMidEndSepPunct{\mcitedefaultmidpunct}
{\mcitedefaultendpunct}{\mcitedefaultseppunct}\relax
\EndOfBibitem
\bibitem[Yu \latin{et~al.}(2010)Yu, Onose, Kanazawa, Park, Han, Matsui,
  Nagaosa, and Tokura]{2010:Yu:Nature}
Yu,~X.~Z.; Onose,~Y.; Kanazawa,~N.; Park,~J.~H.; Han,~J.~H.; Matsui,~Y.;
  Nagaosa,~N.; Tokura,~Y. \emph{Nature (London)} \textbf{2010}, \emph{465},
  901\relax
\mciteBstWouldAddEndPuncttrue
\mciteSetBstMidEndSepPunct{\mcitedefaultmidpunct}
{\mcitedefaultendpunct}{\mcitedefaultseppunct}\relax
\EndOfBibitem
\bibitem[Seki \latin{et~al.}(2012)Seki, Yu, Ishiwata, and
  Tokura]{Tokura_CuOSeO_LTEM_Science_12}
Seki,~S.; Yu,~X.~Z.; Ishiwata,~S.; Tokura,~Y. \emph{Science} \textbf{2012},
  \emph{336}, 198--201\relax
\mciteBstWouldAddEndPuncttrue
\mciteSetBstMidEndSepPunct{\mcitedefaultmidpunct}
{\mcitedefaultendpunct}{\mcitedefaultseppunct}\relax
\EndOfBibitem
\bibitem[Schulz \latin{et~al.}(2012)Schulz, Ritz, Bauer, Halder, Wagner, Franz,
  Pfleiderer, Everschor, Garst, and Rosch]{Rocsh_MnSi_emergent_NatPhys}
Schulz,~T.; Ritz,~R.; Bauer,~A.; Halder,~M.; Wagner,~M.; Franz,~C.;
  Pfleiderer,~C.; Everschor,~K.; Garst,~M.; Rosch,~A. \emph{Nat. Phys.}
  \textbf{2012}, \emph{8}, 301--304\relax
\mciteBstWouldAddEndPuncttrue
\mciteSetBstMidEndSepPunct{\mcitedefaultmidpunct}
{\mcitedefaultendpunct}{\mcitedefaultseppunct}\relax
\EndOfBibitem
\bibitem[Milde \latin{et~al.}(2013)Milde, K\"{o}hler, Seidel, Eng, Bauer,
  Chacon, Kindervater, M\"{u}hlbauer, Pfleiderer, Buhrandt, Sch\"{u}tte, and
  Rosch]{2013:Milde:Science}
Milde,~P.; K\"{o}hler,~D.; Seidel,~J.; Eng,~L.~M.; Bauer,~A.; Chacon,~A.;
  Kindervater,~J.; M\"{u}hlbauer,~S.; Pfleiderer,~C.; Buhrandt,~S.;
  Sch\"{u}tte,~C.; Rosch,~A. \emph{Science} \textbf{2013}, \emph{340},
  1076--1080\relax
\mciteBstWouldAddEndPuncttrue
\mciteSetBstMidEndSepPunct{\mcitedefaultmidpunct}
{\mcitedefaultendpunct}{\mcitedefaultseppunct}\relax
\EndOfBibitem
\bibitem[Fert \latin{et~al.}(2013)Fert, Cros, and
  Sampaio]{2013:Fert:NatureNano}
Fert,~A.; Cros,~V.; Sampaio,~J. \emph{Nat. Nanotech.} \textbf{2013}, \emph{8},
  152--156\relax
\mciteBstWouldAddEndPuncttrue
\mciteSetBstMidEndSepPunct{\mcitedefaultmidpunct}
{\mcitedefaultendpunct}{\mcitedefaultseppunct}\relax
\EndOfBibitem
\bibitem[Nagaosa and Tokura(2013)Nagaosa, and
  Tokura]{Tokura_review_skyrmion_Natnano_13}
Nagaosa,~N.; Tokura,~Y. \emph{Nat. Nanotech.} \textbf{2013}, \emph{8},
  899--911\relax
\mciteBstWouldAddEndPuncttrue
\mciteSetBstMidEndSepPunct{\mcitedefaultmidpunct}
{\mcitedefaultendpunct}{\mcitedefaultseppunct}\relax
\EndOfBibitem
\bibitem[Romming \latin{et~al.}({2013})Romming, Hanneken, Menzel, Bickel,
  Wolter, von Bergmann, Kubetzka, and Wiesendanger]{Romming2013}
Romming,~N.; Hanneken,~C.; Menzel,~M.; Bickel,~J.~E.; Wolter,~B.; von
  Bergmann,~K.; Kubetzka,~A.; Wiesendanger,~R. \emph{{Science}}
  \textbf{{2013}}, \emph{{341}}, {636}\relax
\mciteBstWouldAddEndPuncttrue
\mciteSetBstMidEndSepPunct{\mcitedefaultmidpunct}
{\mcitedefaultendpunct}{\mcitedefaultseppunct}\relax
\EndOfBibitem
\bibitem[Mochizuki \latin{et~al.}(2014)Mochizuki, Yu, Seki, Kanazawa, Koshibae,
  Zang, Mostovoy, Tokura, and Nagaosa]{Tokura_CuOSeO-MnSi_ratchet_Natmater_14}
Mochizuki,~M.; Yu,~X.~Z.; Seki,~S.; Kanazawa,~N.; Koshibae,~W.; Zang,~J.;
  Mostovoy,~M.; Tokura,~Y.; Nagaosa,~N. \emph{Nat. Mater.} \textbf{2014},
  \emph{13}, 241--246\relax
\mciteBstWouldAddEndPuncttrue
\mciteSetBstMidEndSepPunct{\mcitedefaultmidpunct}
{\mcitedefaultendpunct}{\mcitedefaultseppunct}\relax
\EndOfBibitem
\bibitem[Schwarze \latin{et~al.}(2015)Schwarze, Waizner, Garst, Bauer,
  Stasinopoulos, Berger, Rosch, Pfleiderer, and
  Grundler]{2015:Schwarze:NatureMater}
Schwarze,~T.; Waizner,~J.; Garst,~M.; Bauer,~A.; Stasinopoulos,~I.; Berger,~H.;
  Rosch,~A.; Pfleiderer,~C.; Grundler,~D. \emph{Nat. Mater.} \textbf{2015},
  \emph{14}, 478--483\relax
\mciteBstWouldAddEndPuncttrue
\mciteSetBstMidEndSepPunct{\mcitedefaultmidpunct}
{\mcitedefaultendpunct}{\mcitedefaultseppunct}\relax
\EndOfBibitem
\bibitem[M\"{u}nzer \latin{et~al.}(2010)M\"{u}nzer, Neubauer, Adams,
  M\"{u}hlbauer, Franz, Jonietz, Georgii, B\"{o}ni, Pedersen, Schmidt, Rosch,
  and Pfleiderer]{2010:Munzer:PhysRevB}
M\"{u}nzer,~W.; Neubauer,~A.; Adams,~T.; M\"{u}hlbauer,~S.; Franz,~C.;
  Jonietz,~F.; Georgii,~R.; B\"{o}ni,~P.; Pedersen,~B.; Schmidt,~M.; Rosch,~A.;
  Pfleiderer,~C. \emph{Phys. Rev. B} \textbf{2010}, \emph{81}, 041203 (R)\relax
\mciteBstWouldAddEndPuncttrue
\mciteSetBstMidEndSepPunct{\mcitedefaultmidpunct}
{\mcitedefaultendpunct}{\mcitedefaultseppunct}\relax
\EndOfBibitem
\bibitem[Adams \latin{et~al.}(2010)Adams, M\"uhlbauer, Neubauer, M\"unzer,
  Jonietz, Georgii, Pedersen, B\"oni, Rosch, and
  Pfleiderer]{Adams_Domains_2010}
Adams,~T.; M\"uhlbauer,~S.; Neubauer,~A.; M\"unzer,~W.; Jonietz,~F.;
  Georgii,~R.; Pedersen,~B.; B\"oni,~P.; Rosch,~A.; Pfleiderer,~C. \emph{J.
  Phys.: Conf. Ser.} \textbf{2010}, \emph{200}, 032001\relax
\mciteBstWouldAddEndPuncttrue
\mciteSetBstMidEndSepPunct{\mcitedefaultmidpunct}
{\mcitedefaultendpunct}{\mcitedefaultseppunct}\relax
\EndOfBibitem
\bibitem[Yu \latin{et~al.}(2011)Yu, Kanazawa, Onose, Kimoto, Zhang, Ishiwata,
  Matsui, and Tokura]{2011:Yu:NatureMater}
Yu,~X.~Z.; Kanazawa,~N.; Onose,~Y.; Kimoto,~K.; Zhang,~W.~Z.; Ishiwata,~S.;
  Matsui,~Y.; Tokura,~Y. \emph{Nat. Mater.} \textbf{2011}, \emph{10},
  106--109\relax
\mciteBstWouldAddEndPuncttrue
\mciteSetBstMidEndSepPunct{\mcitedefaultmidpunct}
{\mcitedefaultendpunct}{\mcitedefaultseppunct}\relax
\EndOfBibitem
\bibitem[Tonomura \latin{et~al.}({2012})Tonomura, Yu, Yanagisawa, Matsuda,
  Onose, Kanazawa, Park, and Tokura]{2012:Tonomura-MnSi}
Tonomura,~A.; Yu,~X.; Yanagisawa,~K.; Matsuda,~T.; Onose,~Y.; Kanazawa,~N.;
  Park,~H.~S.; Tokura,~Y. \emph{{Nano Lett.}} \textbf{{2012}}, \emph{{12}},
  {1673--1677}\relax
\mciteBstWouldAddEndPuncttrue
\mciteSetBstMidEndSepPunct{\mcitedefaultmidpunct}
{\mcitedefaultendpunct}{\mcitedefaultseppunct}\relax
\EndOfBibitem
\bibitem[Yu \latin{et~al.}({2013})Yu, DeGrave, Hara, Hara, Jin, and
  Tokura]{2013:Yu-MnSi}
Yu,~X.; DeGrave,~J.~P.; Hara,~Y.; Hara,~T.; Jin,~S.; Tokura,~Y. \emph{{Nano
  Lett.}} \textbf{{2013}}, \emph{{13}}, {3755--3759}\relax
\mciteBstWouldAddEndPuncttrue
\mciteSetBstMidEndSepPunct{\mcitedefaultmidpunct}
{\mcitedefaultendpunct}{\mcitedefaultseppunct}\relax
\EndOfBibitem
\bibitem[Parkin \latin{et~al.}({2008})Parkin, Hayashi, and Thomas]{2008:Parkin}
Parkin,~S. S.~P.; Hayashi,~M.; Thomas,~L. \emph{{Science}} \textbf{{2008}},
  \emph{{320}}, {190}\relax
\mciteBstWouldAddEndPuncttrue
\mciteSetBstMidEndSepPunct{\mcitedefaultmidpunct}
{\mcitedefaultendpunct}{\mcitedefaultseppunct}\relax
\EndOfBibitem
\bibitem[Zhang \latin{et~al.}({2015})Zhang, Baker, Komineas, and
  Hesjedal]{2015:Zhang-Topo}
Zhang,~S.; Baker,~A.~A.; Komineas,~S.; Hesjedal,~T. \emph{{Sci. Rep.}}
  \textbf{{2015}}, {15773}\relax
\mciteBstWouldAddEndPuncttrue
\mciteSetBstMidEndSepPunct{\mcitedefaultmidpunct}
{\mcitedefaultendpunct}{\mcitedefaultseppunct}\relax
\EndOfBibitem
\bibitem[Liang \latin{et~al.}({2015})Liang, DeGrave, Stolt, Tokura, and
  Jin]{2015:Liang-MnSi-wires}
Liang,~D.; DeGrave,~J.~P.; Stolt,~M.~J.; Tokura,~Y.; Jin,~S. \emph{{Nat.
  Commun.}} \textbf{{2015}}, \emph{{6}}, {8217}\relax
\mciteBstWouldAddEndPuncttrue
\mciteSetBstMidEndSepPunct{\mcitedefaultmidpunct}
{\mcitedefaultendpunct}{\mcitedefaultseppunct}\relax
\EndOfBibitem
\bibitem[Rajeswaria \latin{et~al.}(2015)Rajeswaria, Pinga, Mancini, Murooka,
  Latychevskaia, McGrouther, Cantoni, Baldini, White, Magrez, Giamarchi,
  R{\o}nnow, and Carbone]{White_Cu2OSeO3_LTEM_PNAS_15}
Rajeswaria,~J.; Pinga,~H.; Mancini,~G.~F.; Murooka,~Y.; Latychevskaia,~T.;
  McGrouther,~D.; Cantoni,~M.; Baldini,~E.; White,~J.~S.; Magrez,~A.;
  Giamarchi,~T.; R{\o}nnow,~H.~M.; Carbone,~F. \emph{Proc. Natl. Acad. Sci.
  U.S.A.} \textbf{2015}, \emph{112}, 14212\relax
\mciteBstWouldAddEndPuncttrue
\mciteSetBstMidEndSepPunct{\mcitedefaultmidpunct}
{\mcitedefaultendpunct}{\mcitedefaultseppunct}\relax
\EndOfBibitem
\bibitem[Bos \latin{et~al.}(2008)Bos, Colin, and
  Palstra]{Thomas_Cu2OSeO3_neutron_PRB_08}
Bos,~J.-W.~G.; Colin,~C.~V.; Palstra,~T. T.~M. \emph{Phys. Rev. B}
  \textbf{2008}, \emph{78}, 094416\relax
\mciteBstWouldAddEndPuncttrue
\mciteSetBstMidEndSepPunct{\mcitedefaultmidpunct}
{\mcitedefaultendpunct}{\mcitedefaultseppunct}\relax
\EndOfBibitem
\bibitem[Seki \latin{et~al.}(2012)Seki, Ishiwata, and
  Tokura]{Tokura_CuOSeO_FE_PRB_12}
Seki,~S.; Ishiwata,~S.; Tokura,~Y. \emph{Phys. Rev. B} \textbf{2012},
  \emph{86}, 060403(R)\relax
\mciteBstWouldAddEndPuncttrue
\mciteSetBstMidEndSepPunct{\mcitedefaultmidpunct}
{\mcitedefaultendpunct}{\mcitedefaultseppunct}\relax
\EndOfBibitem
\bibitem[Yang \latin{et~al.}(2012)Yang, Li, Lu, Whangbo, Wei, Gong, and
  Xiang]{Fudan_Cu2OSeO3_DFT_PRL_12}
Yang,~J.~H.; Li,~Z.~L.; Lu,~X.~Z.; Whangbo,~M.-H.; Wei,~S.-H.; Gong,~X.~G.;
  Xiang,~H.~J. \emph{Phys. Rev. Lett.} \textbf{2012}, \emph{109}, 107203\relax
\mciteBstWouldAddEndPuncttrue
\mciteSetBstMidEndSepPunct{\mcitedefaultmidpunct}
{\mcitedefaultendpunct}{\mcitedefaultseppunct}\relax
\EndOfBibitem
\bibitem[Ruff \latin{et~al.}(2015)Ruff, Lunkenheimer, Loidl, Berger, and
  Krohns]{Berger_Cu2OSeO3_phase_diagram_Srep_15}
Ruff,~E.; Lunkenheimer,~P.; Loidl,~A.; Berger,~H.; Krohns,~S. \emph{Sci. Rep.}
  \textbf{2015}, \emph{5}, 15025\relax
\mciteBstWouldAddEndPuncttrue
\mciteSetBstMidEndSepPunct{\mcitedefaultmidpunct}
{\mcitedefaultendpunct}{\mcitedefaultseppunct}\relax
\EndOfBibitem
\bibitem[White \latin{et~al.}(2012)White, Levati{\'c}, Omrani, Egetenmeyer,
  Pr{\v s}a, {\v Z}ivkovi{\'c}, Gavilano, Kohlbrecher, Bartkowiak, Berger, and
  R{\o}nnow]{White_Cu2OSeO3_E_rotation_IOP_12}
White,~J.~S.; Levati{\'c},~I.; Omrani,~A.~A.; Egetenmeyer,~N.; Pr{\v s}a,~K.;
  {\v Z}ivkovi{\'c},~I.; Gavilano,~J.~L.; Kohlbrecher,~J.; Bartkowiak,~M.;
  Berger,~H.; R{\o}nnow,~H.~M. \emph{J. Phys.: Cond. Matter} \textbf{2012},
  \emph{24}, 432201\relax
\mciteBstWouldAddEndPuncttrue
\mciteSetBstMidEndSepPunct{\mcitedefaultmidpunct}
{\mcitedefaultendpunct}{\mcitedefaultseppunct}\relax
\EndOfBibitem
\bibitem[White \latin{et~al.}(2014)White, Pr\ifmmode~\check{s}\else
  \v{s}\fi{}a, Huang, Omrani, \ifmmode \check{Z}\else
  \v{Z}\fi{}ivkovi\ifmmode~\acute{c}\else \'{c}\fi{}, Bartkowiak, Berger,
  Magrez, Gavilano, Nagy, Zang, and
  R\o{}nnow]{White_Cu2OSeO3_E_rotation_PRL_14}
White,~J.~S.; Pr\ifmmode~\check{s}\else \v{s}\fi{}a,~K.; Huang,~P.;
  Omrani,~A.~A.; \ifmmode \check{Z}\else
  \v{Z}\fi{}ivkovi\ifmmode~\acute{c}\else \'{c}\fi{},~I.; Bartkowiak,~M.;
  Berger,~H.; Magrez,~A.; Gavilano,~J.~L.; Nagy,~G.; Zang,~J.; R\o{}nnow,~H.~M.
  \emph{Phys. Rev. Lett.} \textbf{2014}, \emph{113}, 107203\relax
\mciteBstWouldAddEndPuncttrue
\mciteSetBstMidEndSepPunct{\mcitedefaultmidpunct}
{\mcitedefaultendpunct}{\mcitedefaultseppunct}\relax
\EndOfBibitem
\bibitem[Lin \latin{et~al.}(2014)Lin, Batista, Reichhardt, and
  Saxena]{2014:Lin:PhysRevLett}
Lin,~S.-Z.; Batista,~C.~D.; Reichhardt,~C.; Saxena,~A. \emph{Phys. Rev. Lett.}
  \textbf{2014}, \emph{112}, 187203\relax
\mciteBstWouldAddEndPuncttrue
\mciteSetBstMidEndSepPunct{\mcitedefaultmidpunct}
{\mcitedefaultendpunct}{\mcitedefaultseppunct}\relax
\EndOfBibitem
\bibitem[Effenberger and Pertlik(1985)Effenberger, and
  Pertlik]{1985:Effenberger:MonatshChem}
Effenberger,~H.; Pertlik,~F. \emph{Monatsh. Chem.} \textbf{1985}, \emph{117},
  887\relax
\mciteBstWouldAddEndPuncttrue
\mciteSetBstMidEndSepPunct{\mcitedefaultmidpunct}
{\mcitedefaultendpunct}{\mcitedefaultseppunct}\relax
\EndOfBibitem
\bibitem[Chizhikov and Dmitrienko(2015)Chizhikov, and
  Dmitrienko]{Dmitrienko_Cu2OSeO3_theory_JMMM_15}
Chizhikov,~V.~A.; Dmitrienko,~V.~E. \emph{J. Magn. Magn. Mater.} \textbf{2015},
  \emph{382}, 142--151\relax
\mciteBstWouldAddEndPuncttrue
\mciteSetBstMidEndSepPunct{\mcitedefaultmidpunct}
{\mcitedefaultendpunct}{\mcitedefaultseppunct}\relax
\EndOfBibitem
\bibitem[Kohn(1977)]{1977:Kohn:JPhysSocJpn}
Kohn,~K. \emph{J. Phys. Soc. Jpn} \textbf{1977}, \emph{42}, 2065\relax
\mciteBstWouldAddEndPuncttrue
\mciteSetBstMidEndSepPunct{\mcitedefaultmidpunct}
{\mcitedefaultendpunct}{\mcitedefaultseppunct}\relax
\EndOfBibitem
\bibitem[Belesi \latin{et~al.}(2010)Belesi, Rousochatzakis, Wu, Berger, Shvets,
  Mila, and Ansermet]{Berger_Cu2OSeO3_NMR_PRB_10}
Belesi,~M.; Rousochatzakis,~I.; Wu,~H.~C.; Berger,~H.; Shvets,~I.~V.; Mila,~F.;
  Ansermet,~J.~P. \emph{Phys. Rev. B} \textbf{2010}, \emph{82}, 094422\relax
\mciteBstWouldAddEndPuncttrue
\mciteSetBstMidEndSepPunct{\mcitedefaultmidpunct}
{\mcitedefaultendpunct}{\mcitedefaultseppunct}\relax
\EndOfBibitem
\bibitem[Adams \latin{et~al.}(2012)Adams, Chacon, Wagner, Bauer, Brandl,
  Pedersen, Berger, Lemmens, and Pfleiderer]{Pf_CuOSeO_PRL_12}
Adams,~T.; Chacon,~A.; Wagner,~M.; Bauer,~A.; Brandl,~G.; Pedersen,~B.;
  Berger,~H.; Lemmens,~P.; Pfleiderer,~C. \emph{Phys. Rev. Lett.}
  \textbf{2012}, \emph{108}, 237204\relax
\mciteBstWouldAddEndPuncttrue
\mciteSetBstMidEndSepPunct{\mcitedefaultmidpunct}
{\mcitedefaultendpunct}{\mcitedefaultseppunct}\relax
\EndOfBibitem
\bibitem[Seki \latin{et~al.}(2012)Seki, Kim, Inosov, Georgii, Keimer, Ishiwata,
  and Tokura]{Tokura_CuOSeO_rotation_PRB_12}
Seki,~S.; Kim,~J.-H.; Inosov,~D.~S.; Georgii,~R.; Keimer,~B.; Ishiwata,~S.;
  Tokura,~Y. \emph{Phys. Rev. B} \textbf{2012}, \emph{85}, 220406(R)\relax
\mciteBstWouldAddEndPuncttrue
\mciteSetBstMidEndSepPunct{\mcitedefaultmidpunct}
{\mcitedefaultendpunct}{\mcitedefaultseppunct}\relax
\EndOfBibitem
\bibitem[Adams \latin{et~al.}(2011)Adams, M\"{u}hlbauer, Pfleiderer, Jonietz,
  Bauer, Neubauer, Georgii, B\"{o}ni, Keiderling, Everschor, Garst, and
  Rosch]{2011:Adams:PhysRevLett}
Adams,~T.; M\"{u}hlbauer,~S.; Pfleiderer,~C.; Jonietz,~F.; Bauer,~A.;
  Neubauer,~A.; Georgii,~R.; B\"{o}ni,~P.; Keiderling,~U.; Everschor,~K.;
  Garst,~M.; Rosch,~A. \emph{Phys. Rev. Lett.} \textbf{2011}, \emph{107},
  217206\relax
\mciteBstWouldAddEndPuncttrue
\mciteSetBstMidEndSepPunct{\mcitedefaultmidpunct}
{\mcitedefaultendpunct}{\mcitedefaultseppunct}\relax
\EndOfBibitem
\bibitem[Langner \latin{et~al.}(2014)Langner, Roy, Mishra, Lee, Shi, Hossain,
  Chuang, Seki, Tokura, Kevan, and Schoenlein]{Tokura_CuOSeO_REXS_PRL_14}
Langner,~M.~C.; Roy,~S.; Mishra,~S.~K.; Lee,~J. C.~T.; Shi,~X.~W.;
  Hossain,~M.~A.; Chuang,~Y.~D.; Seki,~S.; Tokura,~Y.; Kevan,~S.~D.;
  Schoenlein,~R.~W. \emph{Phys. Rev. Lett.} \textbf{2014}, \emph{112},
  167202\relax
\mciteBstWouldAddEndPuncttrue
\mciteSetBstMidEndSepPunct{\mcitedefaultmidpunct}
{\mcitedefaultendpunct}{\mcitedefaultseppunct}\relax
\EndOfBibitem
\bibitem[Beale \latin{et~al.}(2010)Beale, Hase, Iida, Endo, Steadman, Marshall,
  Dhesi, van~der Laan, and Hatton]{RASOR_review}
Beale,~T. A.~W.; Hase,~T. P.~A.; Iida,~T.; Endo,~K.; Steadman,~P.;
  Marshall,~A.~R.; Dhesi,~S.~S.; van~der Laan,~G.; Hatton,~P.~D. \emph{Rev.
  Sci. Instrum.} \textbf{2010}, \emph{81}, 073904\relax
\mciteBstWouldAddEndPuncttrue
\mciteSetBstMidEndSepPunct{\mcitedefaultmidpunct}
{\mcitedefaultendpunct}{\mcitedefaultseppunct}\relax
\EndOfBibitem
\bibitem[Hasan and Kane(2010)Hasan, and Kane]{Review_TI}
Hasan,~M.~Z.; Kane,~C.~L. \emph{Rev. Mod. Phys.} \textbf{2010}, \emph{82},
  3045\relax
\mciteBstWouldAddEndPuncttrue
\mciteSetBstMidEndSepPunct{\mcitedefaultmidpunct}
{\mcitedefaultendpunct}{\mcitedefaultseppunct}\relax
\EndOfBibitem
\bibitem[Heinze \latin{et~al.}(2011)Heinze, von Bergmann, Menzel, Brede,
  Kubetzka, Wiesendanger, Bihlmayer, and Bl{\"u}gel]{Hamburg_Ir/Fe}
Heinze,~S.; von Bergmann,~K.; Menzel,~M.; Brede,~J.; Kubetzka,~A.;
  Wiesendanger,~R.; Bihlmayer,~G.; Bl{\"u}gel,~S. \emph{Nat. Phys.}
  \textbf{2011}, \emph{7}, 713--718\relax
\mciteBstWouldAddEndPuncttrue
\mciteSetBstMidEndSepPunct{\mcitedefaultmidpunct}
{\mcitedefaultendpunct}{\mcitedefaultseppunct}\relax
\EndOfBibitem
\bibitem[Karhu \latin{et~al.}(2012)Karhu, R\"o{\ss}ler, Bogdanov, Kahwaji,
  Kirby, Fritzsche, Robertson, Majkrzak, and
  Monchesky]{Monchesky_Karhu_sim_PRB_12}
Karhu,~E.~A.; R\"o{\ss}ler,~U.~K.; Bogdanov,~A.~N.; Kahwaji,~S.; Kirby,~B.~J.;
  Fritzsche,~H.; Robertson,~M.~D.; Majkrzak,~C.~F.; Monchesky,~T.~L.
  \emph{Phys. Rev. B} \textbf{2012}, \emph{85}, 094429\relax
\mciteBstWouldAddEndPuncttrue
\mciteSetBstMidEndSepPunct{\mcitedefaultmidpunct}
{\mcitedefaultendpunct}{\mcitedefaultseppunct}\relax
\EndOfBibitem
\bibitem[Wilson \latin{et~al.}(2013)Wilson, Karhu, Lake, Quigley, Meynell,
  Bogdanov, R{\"o}{\ss}ler, and Monchesky]{Monchesky_Wilson_unwind_PRB_13}
Wilson,~M.~N.; Karhu,~E.~A.; Lake,~D.~P.; Quigley,~A.~S.; Meynell,~S.;
  Bogdanov,~A.~N.; R{\"o}{\ss}ler,~H. F. U.~K.; Monchesky,~T.~L. \emph{Phys.
  Rev. B} \textbf{2013}, \emph{88}, 214420\relax
\mciteBstWouldAddEndPuncttrue
\mciteSetBstMidEndSepPunct{\mcitedefaultmidpunct}
{\mcitedefaultendpunct}{\mcitedefaultseppunct}\relax
\EndOfBibitem
\bibitem[Meynell \latin{et~al.}(2014)Meynell, Wilson, Fritzsche, Bogdanov, and
  Monchesky]{Monchesky_MnSi_surface_twist_PRB_14}
Meynell,~S.~A.; Wilson,~M.~N.; Fritzsche,~H.; Bogdanov,~A.~N.; Monchesky,~T.~L.
  \emph{Phys. Rev. B} \textbf{2014}, \emph{90}, 014406\relax
\mciteBstWouldAddEndPuncttrue
\mciteSetBstMidEndSepPunct{\mcitedefaultmidpunct}
{\mcitedefaultendpunct}{\mcitedefaultseppunct}\relax
\EndOfBibitem
\bibitem[Templeton and Templeton(1994)Templeton, and
  Templeton]{Templeton_PRB_94}
Templeton,~D.~H.; Templeton,~L.~K. \emph{Phys. Rev. B} \textbf{1994},
  \emph{49}, 14850--14853\relax
\mciteBstWouldAddEndPuncttrue
\mciteSetBstMidEndSepPunct{\mcitedefaultmidpunct}
{\mcitedefaultendpunct}{\mcitedefaultseppunct}\relax
\EndOfBibitem
\bibitem[Blume and Gibbs({1988})Blume, and Gibbs]{Blume_1988}
Blume,~M.; Gibbs,~D. \emph{{Phys. Rev. B}} \textbf{{1988}}, \emph{{37}},
  {1779--1789}\relax
\mciteBstWouldAddEndPuncttrue
\mciteSetBstMidEndSepPunct{\mcitedefaultmidpunct}
{\mcitedefaultendpunct}{\mcitedefaultseppunct}\relax
\EndOfBibitem
\bibitem[van~der Laan(2008)]{2008:vdL}
van~der Laan,~G. \emph{C. R. Physique} \textbf{2008}, \emph{9},
  {570--584}\relax
\mciteBstWouldAddEndPuncttrue
\mciteSetBstMidEndSepPunct{\mcitedefaultmidpunct}
{\mcitedefaultendpunct}{\mcitedefaultseppunct}\relax
\EndOfBibitem
\bibitem[R\"o{\ss}ler \latin{et~al.}(2011)R\"o{\ss}ler, Leonov, and
  Bogdanov]{Bogdanov_IOP_2011}
R\"o{\ss}ler,~U.~K.; Leonov,~A.~A.; Bogdanov,~A.~N. \emph{J. Phys.: Conf. Ser.}
  \textbf{2011}, \emph{303}, 012105\relax
\mciteBstWouldAddEndPuncttrue
\mciteSetBstMidEndSepPunct{\mcitedefaultmidpunct}
{\mcitedefaultendpunct}{\mcitedefaultseppunct}\relax
\EndOfBibitem
\end{mcitethebibliography}
\end{document}